\documentclass[smallextended,final,numbook]{svjour3}
\usepackage{amssymb}
\usepackage{amsmath}
\usepackage[latin1]{inputenc}
\usepackage[swedish,english]{babel}
\usepackage{ifpdf}
\ifpdf
  \usepackage[pdftex]{graphicx}
  \usepackage{epstopdf}
\else
  \usepackage[dvips]{graphicx}
\fi
\usepackage{subfigure}
\usepackage{float}
\usepackage[pdftitle={Adaptive FMMs on the GPU},
  pdfauthor={Anders Goude and Stefan Engblom},
  pdffitwindow=true,
  breaklinks=true,
  colorlinks=true,
  urlcolor=blue,
  linkcolor=black,
  citecolor=black,
  anchorcolor=black]{hyperref}
\usepackage{algorithmic}
\usepackage[numbers,sort]{natbib}

\newcommand{\realdom}{\mathbf{R}}
\newcommand{\Ordo}[1]{\mathcal{O}\left(#1\right)}
\newcommand{\thetacriterion}{$\theta$-criterion}
\newcommand{\TOL}{\operatorname{TOL}}

\newtheorem{algorithm}[theorem]{Algorithm}

\begin{document}

\title{Adaptive fast multipole methods on the GPU}

\author{Anders Goude \and Stefan Engblom\thanks{Corresponding author:
    S. Engblom, telephone +46-18-471 27 54, fax +46-18-51 19 25.}}

\institute{A. Goude \at Division of Electricity, Department of
  Engineering Sciences, Uppsala University, SE-751 21 Uppsala, Sweden.
  \\ \email{anders.goude@angstrom.uu.se}.  \and S. Engblom \at
  Division of Scientific Computing, Department of Information
  Technology, Uppsala University, SE-751 05 Uppsala, Sweden.
  \\ \email{stefane@it.uu.se}, \textit{URL:}
  \url{http://user.it.uu.se/~stefane}.}

\date{October 8, 2012}

\maketitle

\begin{abstract}
  We present a highly general implementation of fast multipole methods
  on graphics processing units (GPUs). Our two-dimensional double
  precision code features an asymmetric type of adaptive space
  discretization leading to a particularly elegant and flexible
  implementation. All steps of the multipole algorithm are efficiently
  performed on the GPU, including the initial phase which assembles
  the topological information of the input data. Through careful
  timing experiments we investigate the effects of the various
  peculiarities of the GPU architecture.



\end{abstract}

\selectlanguage{english}


\section{Introduction}


We discuss in this paper implementation and performance issues for
adaptive fast multipole methods (FMMs). Our concerns are focused on
using modern high-throughput graphics processing units (GPUs) which
have seen an increased popularity in Scientific Computing in recent
years. This is mainly thanks to their high peak floating point
performance and memory bandwidth, implying a theoretical performance
which is an order of magnitude better than for CPUs (or even
more). However, in practice for problems in Scientific Computing, the
floating point peak performance can be difficult to realize since many
such problems are bandwidth limited \cite{GPUlimits}. Although the GPU
processor bandwidth is up to 4 times larger than that of the CPU, this
is clearly not sufficient whenever the parallel speedup is (or could
be) much larger than this. Moreover, according to the GPU
computational model, the threads has to be run synchronously in a
highly stringent manner. For these reasons, near optimal performance
can generally only be expected for algorithms of predominantly
data-parallel character.

Another difficulty with many algorithms in Scientific Computing is
that the GPU off-chip bandwidth is comparably small such that the
ability to mask this communication becomes very important
\cite{GPUlimits}. Since the traditional form of many algorithms often
involves intermediate steps for which the GPU architecture is
sub-optimal, a fair degree of rethinking is usually necessary to
obtain an efficient implementation.

Fast multipole methods appeared first in \cite{FMM,AFMM} and have
remained important computational tools for evaluating pairwise
interactions of the type
\begin{align}
  \label{eq:paireval}
  \Phi(x_{i}) &= \sum_{j = 1, j \not = i}^{N} G(x_{i},x_{j}),
  \quad x_{i} \in \realdom^{D}, \quad i = 1 \ldots N,
\end{align}
where $D \in \{2,3\}$. More generally, one may consider to evaluate
\begin{align}
  \label{eq:poteval}
  \Phi(y_{i}) &= \sum_{j = 1, x_{j} \not = y_{i}}^{N} G(y_{i},x_{j}),
  \quad i = 1 \ldots M,
\end{align}
where $\{y_{i}\}$ is a a set of \emph{evaluation points} and
$\{x_{j}\}$ a set of \emph{source points}. In this paper we shall also
conveniently use the terms \emph{potentials} or simply
\emph{particles} to denote the set of sources $\{x_{j}\}$.

Although the direct evaluation of \eqref{eq:paireval} has a complexity
of $\Ordo{N^{2}}$, the task is trivially parallelizable and can be
performed much more efficiently using GPUs than CPUs. For sufficiently
large $N$, however, tree-based codes in general and the FMM algorithm
in particular become important alternatives. The practical complexity
of FMMs scales linearly with the input data and, moreover, effective
\textit{a priori} error estimates are available. Parallel
implementations are, however, often highly complicated and balancing
efficiency with software complexity is not so straightforward
\cite{ace,pace}.

In this paper we present a double precision GPU implementation of the
FMM algorithm which is fully \emph{adaptive}. Although adaptivity
implies a more complex algorithm, this feature is critical in many
important applications. Moreover, in our approach, all steps of the
FMM algorithm are performed on the GPU, thereby reducing memory
traffic to the host CPU.

Successful implementations of the FMM algorithm for GPUs have been
reported previously \cite{fmmgpu,fmmgpu3,exascaleFMM} under certain
limitations. Specifically, with GPUs the performance of single
precision algorithms is a factor of \emph{at least} 2 times better
than double precision \cite[p.~11]{Fermi}. In fact, for
computationally intensive applications, this factor can reach as high
as 8 times \cite{GPUbench}, which implies that single precision
speedups vis-{\`a}-vis CPU implementations can well be $> 2$. It
should be noted, however, that simpler tree-based methods than the FMM
exist that offer a better performance at low tolerance
(cf.~\cite{BarnesHutGPU}, \cite[Chap.~8.7]{griebelMoldyn}), and that
the FMM is of interest mainly for higher accuracy demands.

In Section~\ref{sec:algorithm} we give an overview of our version of
the adaptive FMM. The details of the GPU implementation are found in
Section~\ref{sec:implementation} and in a separate
Section~\ref{sec:differences} we highlight the algorithmic changes
that were made to the original serial code described in
\cite{thetanote}. In Section~\ref{sec:performance} we examine in
detail the speed-ups obtained when moving the various phases of the
algorithm from the CPU to the GPU. We also reason about the results
such that our findings may benefit others who try to port their codes
to the GPU. Since the FMM has been judged to be one of the top 10 most
important algorithms of the 20th century \cite{Top10algorithms}, it is
our hope that insights obtained here is of general value. A final
concluding discussion around these matters is found in
Section~\ref{sec:conclusions}.

\paragraph{Availability of software}

The code discussed in the paper is publicly available and the
performance experiments reported here can be repeated through the
Matlab-scripts we distribute. Refer to Section~\ref{subsec:reproduce}
for details.


\section{Well-separated sets and adaptive multipole algorithms}
\label{sec:algorithm}

In a nutshell, the FMM algorithm is a tree-based algorithm which
produces a continuous representation of the potential field
\eqref{eq:poteval} from all source points in a finite
domain. Initially, all potentials are placed in a single enclosing box
at the zeroth level in the tree. The boxes are then successively split
into child-boxes such that the number of points per box decreases with
each level.

\begin{figure}
  \subfigure[]{\includegraphics{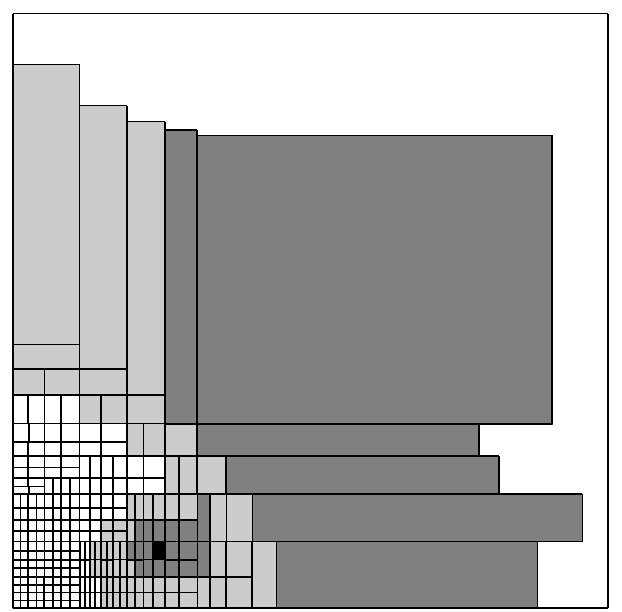}}
  \subfigure[]{\includegraphics{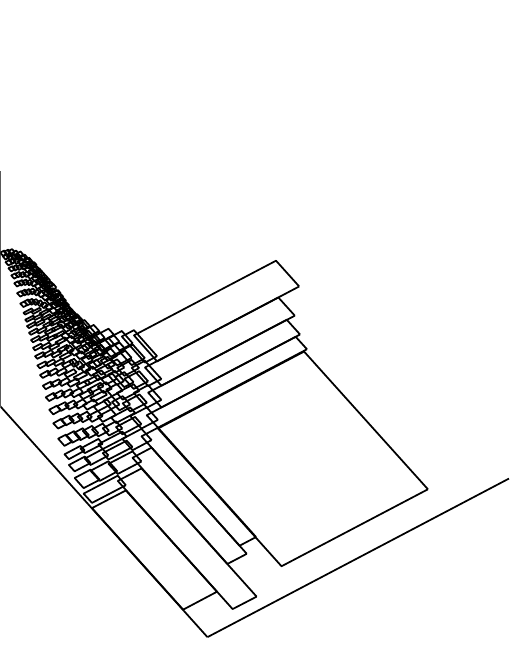}}
  \caption{The adaptive mesh is constructed by recursively splitting
    boxes along the coordinate axes in such a way that the number of
    source points is very nearly the same in the four resulting
    boxes. \textit{(a)} Here the boxes colored in light gray will
    interact via multipole-to-local shifts with the black box, that
    is, they satisfy the \thetacriterion\ ($\theta = 1/2$). The boxes
    in dark gray are strongly connected to the black box and must be
    taken care of at the next level in the tree. \textit{(b)} Same
    mesh as in \textit{(a)}, but visualized as a distribution by
    letting the height of each box be inversely proportional to its
    area. The source points in this example were sampled from a normal
    distribution.}
  \label{fig:mesh}
\end{figure}

The version of FMM that we consider is described in \cite{thetanote}
and is organized around the requirement that boxes at the same level
in the tree are either \emph{decoupled} or \emph{strongly/weakly
  coupled}. The type of coupling between the boxes follows from the
\emph{\thetacriterion}, which states that for two boxes with radii
$r_{1}$ and $r_{2}$, whose centers are separated with distance $d$,
the boxes are \emph{well-separated} whenever
\begin{align}
  \label{eq:thetacriterion}
  R+\theta r \le\theta d,
\end{align}
where $R = \max\left\{r_{1},r_{2}\right\}$, $r = \min\left\{
r_{1},r_{2}\right\}$, and $\theta \in (0,1)$ a parameter. In this
paper we use the constant value $\theta = 1/2$ which we have found to
perform well in practice. At each level $l$ and for each box $b$, the
set $S(p)$ of strongly coupled boxes of its parent box $p$ is
examined; children of $S(p)$ that satisfy the \thetacriterion\ with
respect to $b$ are allowed to become weakly coupled to $b$, otherwise
they remain strongly coupled. Since a box is defined to be strongly
connected to itself this rule defines the connectivity for the whole
multipole tree. In Figure~\ref{fig:mesh} an example of a multipole
mesh and its associated connectivity pattern is displayed.

All boxes are equipped with an outgoing \emph{multipole} expansion and
an incoming \emph{local} expansion. The multipole expansion is the
expansion of all sources within the box around its center and is valid
away from the box. To be concrete, in our two-dimensional
implementation we use the traditional format of a $p$-term expansion
in the complex plane,
\begin{align}
  \label{eq:multipole}
  M(z) &= a_{0} \log(z-z_{0})+\sum_{j = 1}^{p} \frac{a_{j}}{(z-z_{0})^{j}},
\end{align}
where $z_{0}$ is the center of the box. The local expansion is instead
the expansion of sources far away from the box and can therefore be
used to evaluate the contribution from these sources at all points
within the box:
\begin{align}
  \label{eq:local}
  L(z) &= \sum_{j = 0}^{p} b_{j} (z-z_{0})^{j},
\end{align}
where again \eqref{eq:local} is specific to a two-dimensional
implementation.

The computational part of the FMM algorithm proceeds in an
\emph{upward} and a \emph{downward} phase. During the first stage the
\emph{multipole-to-multipole} (M2M) shift from children to parent
boxes recursively propagates and accumulates multipole expansions. In
the second stage the important \emph{multipole-to-local} (M2L) shift
adds to the local expansions in all weakly coupled boxes which are
then propagated downwards to children through the
\emph{local-to-local} shift (L2L). At the finest level, any remaining
strong connections are evaluated through direct evaluation of
\eqref{eq:paireval} or \eqref{eq:poteval}. A simple optimization which
was noted already in \cite{AFMM} is that, at the lowest level,
strongly coupled boxes are checked with respect to the
\thetacriterion~\eqref{eq:thetacriterion}, but \emph{with the roles of
  $r$ and $R$ interchanged}. If found to be true, then the potentials
in the larger box can be directly shifted into a local expansion in
the smaller box, and the outgoing multipole expansion from the smaller
can directly be evaluated within the larger box.

The algorithm so far has been described without particular assumptions
on the multipole mesh itself. As noted in \cite{thetanote}, relying on
\emph{asymmetric adaptivity} when constructing the meshes makes a very
convenient implementation possible. In particular, this construction
avoids complicated cross-level communication and implies that the
multipole tree is \emph{balanced}, rendering the use of post-balancing
algorithms \cite{balancedtrees} unnecessary. Also, the possibility to
use a static layout of memory is particularly attractive when
considering data-parallel implementations for which the benefits with
adaptive meshes have been questioned \cite{fmmgpu}.

In this scheme, the child boxes are created by successively splitting
the parent boxes close to the median value of the particle positions,
causing all child boxes to have about the same number of particles. At
each level, all boxes are split twice in succession, thus producing
four times as many boxes for each new level. The resulting FMM tree is
a \emph{pyramid} rather than a general tree for which the depth may
vary. The cost is that with a balanced tree, the communication stencil
is variable. Additionally, it also prevents the use of certain
symmetries in the multipole translations as described in
\cite{Hrycak98}. To improve on the communication locality, the
direction of the split is guided by the eccentricity of the box since
the algorithm gains in efficiency when the boxes have equal width and
height (the \thetacriterion\ is rotationally invariant).

The \emph{algorithmic complexity} of the FMM has been discussed by
many authors. Practical experiences \cite{nbodycompare},
\cite[Chap.~8.7]{griebelMoldyn}, \cite[Chap.~6.6.3]{helmholtzFMM},
indicate that linear complexity in the number of source points is
observed in most cases, but that simpler algorithms perform better in
certain situations. Although it is possible to construct explicit
distributions of points for which the FMM algorithm has a
\emph{quadratic complexity} \cite{FMMcomplexitynote}, this behavior is
usually not observed in practical applications.

With $p$ terms used in both the multipole and the local expansions, we
expect the serial computational complexity of our implementation to be
proportional to $\theta^{-2} p^{2} \cdot N$, with $N$ the number of
source points. This follows from assuming an asymptotically regular
mesh such that $R \sim r$ in \eqref{eq:thetacriterion} and a total of
on the order of $N$ boxes at the finest level. Then each of those $N$
boxes interact through M2L-interactions with about $\pi d^{2} \times
N$ other boxes. From \eqref{eq:thetacriterion} we get $d \sim
(1+\theta)/\theta \times r \sim (\sqrt{N} \theta)^{-1}$, and since the
M2L-interaction is a linear mapping between $p$ coefficients this
explains our stated estimate. This simple derivation assumes that the
M2L-shift is the most expensive part of the algorithm. In practice,
the cost of the direct evaluation of \eqref{eq:paireval} may well
match this part. From extensive experiments with the serial version of
the algorithm, we have seen that it is usually possible to balance the
algorithm in such a way that these two parts take roughly the same
time.

With a given relative target tolerance $\TOL$, the analysis in
\cite{thetanote} implies $p \sim \log\TOL/\log\theta$, so that the
total complexity can be expected to be on the order of $\theta^{-2}
\log^{-2}\theta \cdot N\log^{2} \TOL$. We now proceed to discuss an
implementation which distributes this work very efficiently on a GPU
architecture.


\section{GPU Implementation}
\label{sec:implementation}

The first major part of the adaptive fast multipole method is the
\emph{topological phase} which arranges the input into a hierarchical
\emph{FMM mesh} and determines the type of interactions to be
performed. We discuss this part in Section~\ref{subsec:sorting}. The
second part is discussed in Section~\ref{subsec:multipole} and
consists of the actual multipole evaluation algorithm with its upward
and downward phases performing the required interactions in a
systematic fashion.

To motivate some of our design choices, we choose to start with a
brief discussion on the GPU hardware and the CUDA programming model
\emph{(Compute Unified Device Architecture)}. The interested reader is
referred to \cite{CUDA42, Farber11} for further details.

\subsection{Overview of the GPU architecture}

Using CUDA terminology, the GPU consists of several
\emph{multiprocessors} (14 for the Tesla C2075), where each
multiprocessor contains many CUDA \emph{cores} (32 for the C2075). The
CUDA execution model groups 32 threads together into a
\emph{warp}. All threads in the same warp execute the same
instruction, but operate on different data; this is simply the
GPU-version of the \emph{Single Instruction Multiple Data
  (SIMD)}-model for parallel execution. Whenever two threads within
the same warp need to execute different instructions, the operations
become serialized.

To perform a calculation on the GPU, the CPU launches a \emph{kernel}
containing the code for the actual calculation. The threads on the GPU
are grouped together into \emph{blocks}, where the number of blocks as
well as the number of threads per block is to be specified at each
kernel launch (for performance reasons the number of threads per block
is usually a multiple of the warp size). All threads within a block
will be executed on the same multiprocessor and thread synchronization
can only be performed efficiently between the threads of a single
block. This synchronization is required whenever two threads write to
the same memory address. Although for this purpose there are certain
built-in \emph{atomic write} operations, none of these support double
precision for the Tesla C2075.

The GPU typically has access to its own global memory and all data has
to be copied to this memory before being used. Further, each
multiprocessor has a special fast \emph{shared memory}, which can be
used for the communication between threads within the same block
\cite[Chap.~5]{Farber11}. For example, the C2075 has 48 kB of shared
memory per multiprocessor for this purpose. Overuse of this memory
limits the number of threads that can run simultaneously on a
multiprocessor, which in turn has a negative effect on performance.

\subsection{Topological phase}
\label{subsec:sorting}

The topological phase consists of two parts, where the first part
creates the boxes by partitioning the particles (we refer to this as
\emph{``sorting''}) and the second part determines the interactions
between them (\emph{``connecting''}).

The sorting algorithm successively partitions each box in two parts
according to a chosen pivot point
(Algorithm~\ref{alg:Partitioning-with-successive}). The pivot element
is obtained by first sorting 32 of the elements using a simple
$\Ordo{n^{2}}$ algorithm where each thread sorts a single point (32
elements was chosen to match the warp size). The pivot is then
determined by interpolation of the current relative position in the
active set of points so as to approximately land at the global median
point (line~\ref{line:pivot_32},
Algorithm~\ref{alg:Partitioning-with-successive}).

\begin{algorithm}[Partitioning with successive splits]
  \label{alg:Partitioning-with-successive}
  \begin{algorithmic}
    \STATE
    \STATE \textbf{Input:} \textnormal{Unordered array consisting
    of $x$- or $y$-coordinates.}
    \STATE \textbf{Output:} \textnormal{Array partitioned around its median
    coordinate.}
  \end{algorithmic}
  \begin{algorithmic}[1]  
    \WHILE{size(array) $>$ 32}
    \STATE determine\_pivot\_32()\label{line:pivot_32}
    \STATE split\_around\_pivot()\label{line:split_pivot}
    \STATE keep\_part\_containing\_median()\label{line:keep_part}
    \ENDWHILE
    \STATE
    \COMMENT{the array now consists of $\le$ 32 elements:}
    \STATE determine\_median\_32()\label{line:median_32}
  \end{algorithmic}
\end{algorithm}

The split in line~\ref{line:split_pivot} uses a two-pass scheme, where
each thread handles a small set of points to split. The threads start
by counting the number of points smaller than the pivot. Then a global
cumulative sum has to be calculated. Within a block, the method
described in \cite{harris_07} is used. For multiple blocks,
\emph{atomic addition} is used in between the blocks, thus allowing
the split to be performed in a single kernel call (note that using
atomic addition makes the code non-deterministic). Given the final
cumulative sum, the threads can correctly insert their elements in the
sorted array (second pass). After the split, only the part containing
the median is kept for the next pass (line~\ref{line:keep_part}).

Algorithm~\ref{alg:Partitioning-with-successive} can be used to
partition many boxes in a single kernel call using one block per
box. If the number of boxes is low, it is desirable to use several
blocks for each partitioning to better use the GPU cores. This
requires communication between the blocks and the partitioning has to
be performed with several kernel calls according to
Algorithm~\ref{alg:Partitioning-multiblock}. The splitting code is
executed in a loop (lines \ref{line:partition_while_start}
to~\ref{line:partition_while_end}) and a small amount of data transfer
between the GPU and the CPU is required to determine the number of
loops.

\begin{algorithm}[Partitioning with successive splits, CPU part]
  \label{alg:Partitioning-multiblock}
  \begin{algorithmic}
    \STATE
    \STATE \textbf{Input:} \textnormal{Unordered array consisting
    of $x$- or $y$-coordinates.}
    \STATE \textbf{Output:} \textnormal{Array partitioned around its median
    coordinate.}
  \end{algorithmic}
  \begin{algorithmic}[1]  
    \STATE determine\_split\_direction()
    \WHILE {$\max_{i}size(array_{i}) > $ single\_thread\_limit}\label{line:partition_while_start}
    \STATE \COMMENT{executed in parallel:}
    \FORALL{splits}
    \STATE determine\_pivot\_32()
    \STATE partition\_around\_pivot()
    \STATE keep\_part\_containing\_median()
    \ENDFOR
    \ENDWHILE\label{line:partition_while_end}
    \STATE split\_on\_single\_block()\label{line:single_block_split}
  \end{algorithmic}
\end{algorithm}

Running the code using multiple blocks forces the code to run
synchronized, with equal amount of splits in each partitioning. If one
bad pivot is encountered, then this split takes much longer time than
the others resulting in bad parallel efficiency. By contrast, a
\emph{single} block running does not force this synchronization and,
additionally, allows for a better caching of the elements since they
remain in the same kernel call all the time. For these reasons,
Algorithm~\ref{alg:Partitioning-multiblock} switches to single block
mode (line~\ref{line:single_block_split}) when all splits contain a
small enough number of points ($single\_thread\_limit=4096$ in the
current implementation).

The second part of the topological phase determines the
\emph{connectivity} of the FMM mesh, that is, if the boxes should
interact via near- or far-field interactions. This search is performed
for each box independently and the parallelization one thread/box is
used here. Each kernel call calculates the interactions of one full
level of the FMM tree.

\subsection{Computational part: the multipole algorithm}
\label{subsec:multipole}

The computational part consists of all the multipole-related
interactions, which include initialization (P2M), shift operators
(M2M), (M2L), and (L2L), and local evaluation (L2P). Additionally, we
also include the direct interaction in the near-field (P2P) in this
floating point intensive phase of the FMM algorithm. During the
computational part, no data transfer is necessary between the GPU and
the host.

\subsubsection{Multipole initialization}

The initialization phase creates multipole expansions for each box via
\emph{particle-to-multipole} shifts (P2M). Since each source point
gives a contribution to each coefficient $a_j$, using several threads
per box requires intra-thread communication which in
Algorithm~\ref{alg:Multipole-initialization} is accomplished by
introducing a temporary matrix to store the coefficients.

Initially (line~\ref{line:calc_Ncache},
Algorithm~\ref{alg:Multipole-initialization}), one thread calculates
the coefficients for one source particle (two threads if the number of
particles is less than half the number of available threads). Then
each thread calculates the sum for one coefficient
(line~\ref{line:sum_array}). This procedure has to be repeated for a
large number of coefficients, as it is desirable to have a small
temporary matrix to limit the use of shared memory. The current
implementation uses 64 threads per box (two warps) and takes 4
coefficients in each loop iteration (8 in the two threads/particle
case).

The initialization also handles the special case where the particles
are converted directly to local expansions via
\emph{particle-to-local} expansions (P2L). The principle for creating
local expansions is the same as for the multipole expansions. All
timings of this phase will include both P2M and P2L shifts.

\begin{algorithm}[Multipole initialization]
  \label{alg:Multipole-initialization}
  \begin{algorithmic}
    \STATE
    \STATE \textbf{Input:} \textnormal{Positions and strengths for
      source particles in a box.}
    \STATE \textbf{Output:} \textnormal{Multipole coefficients $a_i$
      for the box.}
  \end{algorithmic}
  \begin{algorithmic}[1]  
    \STATE \COMMENT{executed in parallel:}
    \FORALL{sources in box}
    \STATE load\_one\_source\_per\_thread()
    \FOR{$k=1$ \textbf{to} $p$}
    \STATE temp\_array := calc\_N\_cache\_coefficients()\label{line:calc_Ncache}
    \STATE synchronize\_threads()
    \STATE $a_k$ := $a_k$ + sum(temp\_array)\label{line:sum_array}
    \ENDFOR
    \ENDFOR
  \end{algorithmic}
\end{algorithm}

\subsubsection{Upward pass}

In the upward pass, the coefficients of each of four child boxes are
repeatedly shifted to their parent's coefficients via the
M2M-shift. This is achieved using Algorithm~\ref{alg:Upward-pass}~(a)
which is similar to the one proposed in
\cite{liu2009fast}. Algorithm~\ref{alg:Upward-pass}~(a) can be
parallelized by allowing one thread to calculate one interaction, and
at the end compute the sum over the four boxes
(line~\ref{line:sum_translations},
Algorithm~\ref{alg:Upward-pass}). It should be noted that the
multiplication in Algorithm~\ref{alg:Upward-pass} is a complex
multiplication which is performed $\Ordo{p^{2}}$ times. By introducing
scaling, the algorithm can be modified to
Algorithm~\ref{alg:Upward-pass}~(b), which instead requires one
complex division, $\Ordo{p}$ complex multiplications and
$\Ordo{p^{2}}$ complex additions. The advantage of this modification
in the GPU case is not the reduction of complex multiplications, but
rather that the real and imaginary parts are independent. Hence two
threads per shift can be used thus reducing the amount of shared
memory per thread.

\begin{algorithm}[Multipole to multipole translation]
  \label{alg:Upward-pass}
  \begin{algorithmic}
    \STATE
    \STATE \textbf{Input:} \textnormal{Multipole coefficients $a_j$ of
      child box at position $z_c$.}
    \STATE \textbf{Output:} \textnormal{Multipole coefficients $a_j$
      of parent box at position $z_p$.}
  \end{algorithmic}
  \vspace{1mm}
  \noindent
  \begin{minipage}[b]{0.48\textwidth}
  \noindent
  \mbox{\textnormal{(a)~Without~scaling}}
  \vspace{-4.5mm}
  \begin{algorithmic}[1]
    \STATE $r:=z_c-z_p$
    \STATE
    \STATE
    \STATE
    \FOR{$k=p$ \textbf{downto} 2}
    \FOR{$j=k$ \textbf{to} $p$}
    \STATE $a_j:=a_j+r \cdot a_{j-1}$
    \ENDFOR
    \ENDFOR
    \FOR{$j=1$ \textbf{to} $p$}
    \STATE $a_j:=a_j-r^j \cdot a_{0}/j$
    \ENDFOR
    \STATE \COMMENT{4 child boxes shift coefficients to the same parent:}
    \STATE $a:=$ sum\_translations()
  \end{algorithmic}
  \end{minipage}
  \noindent
  \begin{minipage}[b]{0.48\textwidth}
  \noindent
  \mbox{\textnormal{(b)~With~scaling}}
  \vspace{-4.5mm}
  \begin{algorithmic}
    \STATE $r:=z_c-z_p$
    \FOR{$j=1$ \textbf{to} $p$}
    \STATE $a_j:=a_j/r^j$
    \ENDFOR
    \FOR{$k=p$ \textbf{downto} 2}
    \FOR{$j=k$ \textbf{to} $p$}
    \STATE $a_j:=a_j+a_{j-1}$
    \ENDFOR
    \ENDFOR
    \FOR{$j=1$ \textbf{to} $p$}
    \STATE $a_j:=(a_j-a_0/j) \cdot r^j$
    \ENDFOR
    \STATE
    \STATE \label{line:sum_translations}
    \STATE $a:=$ sum\_translations()
  \end{algorithmic}
  \end{minipage}
\end{algorithm}

\subsubsection{Downward pass}

The downward pass consists of two parts, the translation of multipole
expansions to local expansions (M2L), and the translation of local
expansions to the children of a box (L2L). The translation of local
expansions to the children is very similar to the M2M-shift discussed
previously, and can be achieved with the scheme in
Algorithm~\ref{alg:Polonomial-translation}~(b). This shift is slightly
simpler on the GPU since there is no need to sum the coefficients at
the end, but instead requires more memory accesses as the calculated
local coefficients must be added to already existing values.

\begin{algorithm}[Local to local translation]
  \label{alg:Polonomial-translation}
  \begin{algorithmic}
    \STATE
    \STATE \textbf{Input:} \textnormal{Local coefficients $b_j$ of
      parent box at position $z_p$.}
    \STATE \textbf{Output:} \textnormal{Local coefficients $b_j$ of
      child box at position $z_c$.}
  \end{algorithmic}
  \begin{algorithmic}[1]
    \STATE $r:=z_p-z_c$
    \FOR{$j=1$ \textbf{to} $p$}
    \STATE $b_j:=b_j \cdot r^j$
    \ENDFOR
    \FOR{$k=0$ \textbf{to} $p$}
    \FOR{$j=p-k$ \textbf{to} $p-1$}
    \STATE $b_j:=b_j-b_{j+1}$
    \ENDFOR
    \ENDFOR
    \FOR{$j=1$ \textbf{to} $p$}
    \STATE $b_j:=b_j/r^j$
    \ENDFOR
  \end{algorithmic}
\end{algorithm}

The translation of multipole expansions to local expansions is the
most time consuming part of the downward pass. The individual shifts
can be performed with a combination of the reduction scheme in the M2M
translation and the L2L translation, see
Algorithm~\ref{alg:Multipole-to-polynomial}. Again, this
implementation allows for two dedicated threads for each shift. We
have not seen this algorithm described elsewhere.

\begin{algorithm}[Multipole to local translation]
  \label{alg:Multipole-to-polynomial}
  \begin{algorithmic}
    \STATE
    \STATE \textbf{Input:} \textnormal{Multipole coefficients $a_j$ of
      box at position $z_i$.}
    \STATE \textbf{Output:} \textnormal{Local coefficients $b_j$ of
      box at position $z_o$.}
  \end{algorithmic}
  \begin{algorithmic}[1]
    \STATE $r:=z_o-z_i$
    \FOR{$j=1$ \textbf{to} $p$} \label{line:m2l_prescale_start}
    \STATE $b_{j-1}:=a_j/r^j \cdot (-1)^j$
    \ENDFOR \label{line:m2l_prescale_end}
    \STATE $b_p:=0$
    \FOR{$k=2$ \textbf{to} $p$}
    \FOR{$j=p-k$ \textbf{to} $p-1$}
    \STATE $b_j:=b_j-b_{j+1}$
    \ENDFOR
    \ENDFOR
    \FOR{$k=p$ \textbf{downto} 1}
    \FOR{$j=k$ \textbf{to} $p$}
    \STATE $b_j:=b_j+b_{j-1}$
    \ENDFOR
    \ENDFOR
    \STATE $b_0:=b_0-a_0 \cdot \log (r)$
    \FOR{$j=1$ \textbf{to} $p$} \label{line:m2l_postscale_start}
    \STATE $b_j:=(b_j-a_0/j)/r^j$
    \ENDFOR \label{line:m2l_postscale_end}
    \STATE $b:=$ sum\_translations()
  \end{algorithmic}
\end{algorithm}

With the chosen adaptive scheme, the number of shifts per box
varies. Since our GPU does not support atomic addition in double
precision, one block has to handle all shifts of one box in order to
perform this operation in one kernel call. As the number of
translations is not always a multiple of 16 (the number of
translations per loop if 32 threads/block is used), this can result in
idle threads. One can partially compensate for this by giving one
block the ability to operate on two boxes in the same loop.

As the M2L translations are performed within a single level of the
multipole tree, all translations can be performed in a single kernel
call. This is in contrast to the M2M- and L2L translations, which both
require one kernel call per level.


\subsubsection{Local evaluation}

The local evaluation (L2P) is scheduled in parallel by using one block
to calculate the interactions of one box. Moreover, one thread
calculates the interactions of one evaluation point from the local
coefficients of the box. This operation requires no thread
communication and can be performed in the same way as for the CPU. The
local evaluation uses 64 threads/block.

This phase will also include the special case where the evaluation is
performed directly through the multipole expansion (M2P). This
operation is performed in a similar way as for the L2P evaluation. All
timings of this phase therefore include both L2P- and M2P evaluations.

\subsubsection{Near-field evaluation}

In the near-field evaluation (P2P), the contribution $F$ from all
boxes within the near-field of a box should be calculated at all
evaluation points of the box. Similar to the M2L translations, the
number of boxes in the near-field varies due to the adaptivity.

\begin{algorithm}[Direct evaluation between boxes]
  \label{alg:Direct-evaluation-box}
    \begin{algorithmic}
    \STATE
    \STATE \textbf{Input:} \textnormal{Positions and strengths of
      particles in the near field.}
    \STATE \textbf{Output:} \textnormal{The contribution $F$ of
      the particles in the near field.}
  \end{algorithmic}
  \begin{algorithmic}[1]
    \STATE \COMMENT{executed in parallel:}
    \STATE load\_evaluation\_point\_positions() \COMMENT{one per thread}
    \FORALL{interaction boxes}
    \STATE cache\_interaction\_positions()\label{line:load_points}
    \IF{ cache\_is\_full \textbf{or} all\_positions\_loaded}
    \FORALL{elements in cache}
    \STATE $F$ := $F$ + add\_pairwise\_interaction()\label{line:interaction}
    \ENDFOR
    \ENDIF
    \ENDFOR
  \end{algorithmic}
\end{algorithm}

The GPU implementation uses one block per box and one to four threads
per evaluation point (depending on the number of evaluation points
compared to the number of threads). The interaction is calculated
according to Algorithm~\ref{alg:Direct-evaluation-box}, where the
source points are loaded into a cache in shared memory
(line~\ref{line:load_points}) and when the cache is full, the
interactions are calculated (line~\ref{line:interaction}). This part
uses 64 threads per block and a suitable cache size is to use the same
size as the number of threads.

\section{Differences between the CPU and GPU implementations}
\label{sec:differences}

In this section we highlight the main differences between the two
versions of the code. A point to note is that when we compare speed-up
with respect to the CPU-code in Section~\ref{sec:performance}, we have
taken care in implementing several optimizations which are CPU-specific.

\subsection{Topological phase}

When sorting, the GPU implementation is based on sorting 32 element
arrays for choosing a pivot element. This design was made to achieve a
better use of the CUDA cores and, in the multiple block/partitioning
case, to make all partitionings behave more similar to each other. The
single-threaded CPU version uses the \emph{median-of-three} procedure
for choosing the pivot element, which is often used in the well-known
algorithm \textit{quicksort} \cite[Chap.~9]{AlgorithmsC}. An advantage
with this method compared to the GPU algorithm is that it is in-place
and hence that there is no need for temporary storage.

\subsection{Computational part}

The direct evaluation can use symmetry on the CPU if the evaluation
points are the same as the source points since the direct interaction
is symmetric. Using this symmetry, the run time of the direct
interaction can be reduced by almost a factor of two in the CPU
version. This is not implemented on the GPU as it would require atomic
memory operations to store the results (which is not available for
double precision on our GPU).

The operations M2M, P2P, P2M, and M2P are all in principle the same in
both versions of the code. For the M2L shift, the symmetry of the
\thetacriterion\ \eqref{eq:thetacriterion} can be used in the scaling
phases on the CPU
(lines~\ref{line:m2l_prescale_start}--\ref{line:m2l_prescale_end} and
\ref{line:m2l_postscale_start}--\ref{line:m2l_postscale_end} in
Algorithm~\ref{alg:Multipole-to-polynomial}), while in the GPU
version, the two shifts are handled by different blocks, making
communication unpractical. Another difference is that the scaling
vector is saved in memory from the pre-scaling part to the
post-scaling part after the principal shift operator. This was
intentionally omitted from the GPU version as the extra use of shared
memory decreased the number of active blocks on each multiprocessor,
and this in turn reduced performance.

\subsection{Symmetry and connectivity}

The CPU implementation uses symmetry throughout the multipole
algorithm.  With symmetry, it is only required to create
one-directional interaction lists when determining the connectivity.

As the GPU implementation does not rely on symmetry when evaluating,
it is beneficial to create directed interaction lists. This causes
twice the work and twice the memory usage (for the connectivity
lists).  However, the time required to determine the connectivity is
quite small ($\sim$ 1\%, see Table~\ref{tab:pie}).

\subsection{Further notes on the CPU implementation}

The current CPU implementation is a single threaded version. Other
research \cite{PetFMM,GPUlimits} has shown that good parallel
efficiency can be obtained for a multicore version (e.g.~85\% parallel
efficiency on 64 processors). However, the work in \cite{PetFMM} does
not appear to use the symmetry in the compared direct evaluation.

In order to achieve a highly efficient CPU implementation, as
suggested also by others \cite{exascaleFMM,SIMD_Nbody}, the multipole
evaluation part was written using SSE intrinsics. Using these
constructs means that two double precision operations can be performed
as one single operation. The direct- and multipole evaluation, as well
as the multipole initialization all use this optimization, where two
points are thus handled simultaneously. Additionally, all shift
operators also rely on SSE intrinsics in their implementation.

\subsection{Double versus single precision}

The entire algorithm is written in double precision for both the CPU
and the GPU. Using single precision would be significantly faster,
both on the CPU (as SSE could operate on 4 floats instead of 2
doubles) and on the GPU (where the performance difference appears to
vary depending on the mathematical expression \cite{GPUbench}). It is
likely that higher speedups could be obtained with a single precision
code, but it would also seriously limit the applicability of the
algorithm. With our approach identical accuracy is obtained from the
two codes.


\section{Performance}
\label{sec:performance}

This section compares the CPU and GPU codes performance-wise. The
simulations were performed on a Linux Ubuntu 10.10-system with an
Intel Xeon 6c W3680 3.33 GHz CPU and a Nvidia Tesla C2075 GPU. The
compilers were GCC 4.4.5 and CUDA 4.0. For all comparisons of
individual parts, the sorting was performed on the CPU to ensure
identical multipole trees (the GPU sorting algorithm is
non-deterministic). All timings have been handled using the timing
functionality of the GPU. The total time includes the time to copy
data between the host and the GPU, while the time of the individual
parts does not include this. All simulations were performed a large
number of times such that the error in the mean when regarded as an
estimator to the expected value was negligible. Overall we found that
the measured times displayed a surprisingly small spread with usually
a standard deviation which was only some small fraction of the
measured value itself.

All performance experiments were conducted using the \emph{harmonic}
potential,
\begin{align}
  G(z_{i},z_{j}) &\equiv \frac{\Gamma_{j}}{z_{j}-z_{i}},
\end{align}
in \eqref{eq:paireval} and hence $a_{0} = 0$ in
\eqref{eq:multipole}. Moreover, in Sections~\ref{subsec:calibration}
through \ref{subsec:breakeven}, all simulations were performed using
randomly chosen source points, homogeneously distributed in the unit
square.

We remark again that the performance experiments reported here can be
repeated using the scripts distributed with the code itself, see
Section~\ref{subsec:reproduce}.

\subsection{Calibration}
\label{subsec:calibration}

From the perspective of performance, the most critical parameter is
the number of levels in the multipole tree. Adding one extra level
increases the number of boxes at the finest level in the tree by
four. Assuming that each box connects to approximately the same number
of boxes at each level, the total number of pairwise interactions
therefore decrease with a factor of about 4. The initialization and
multipole evaluation require the same amount of operations, but will
operate on an increasing number of boxes, thus increasing the memory
accesses. For all shift operations, one additional level implies about
a threefold increase of the total number of interactions and the same
applies for the determination of the connectivity information. For the
sorting, each level requires about the same amount of work, but
handling many small boxes easily causes additional overhead.

It is expected that the CPU code will follow this scaling quite well,
while for the GPU, where several threads should run synchronously,
this is certainly not always the case. As an example, L2P operates by
letting one thread handle the interaction of one source point, P2M can
use up to two threads, and P2P can use up to four threads/point (all
these use 64 threads/block). On the tested Tesla C2075 system, this
means that the local evaluation of a box containing 1 evaluation point
takes the same amount of time as a box containing 64 evaluation points
(on a Geforce GTX 480 system, this only applies to up to 32 evaluation
points which is the warp size here). This shows the sensitivity of the
GPU implementation with respect to the number of points in each box.

\begin{figure}
  \includegraphics{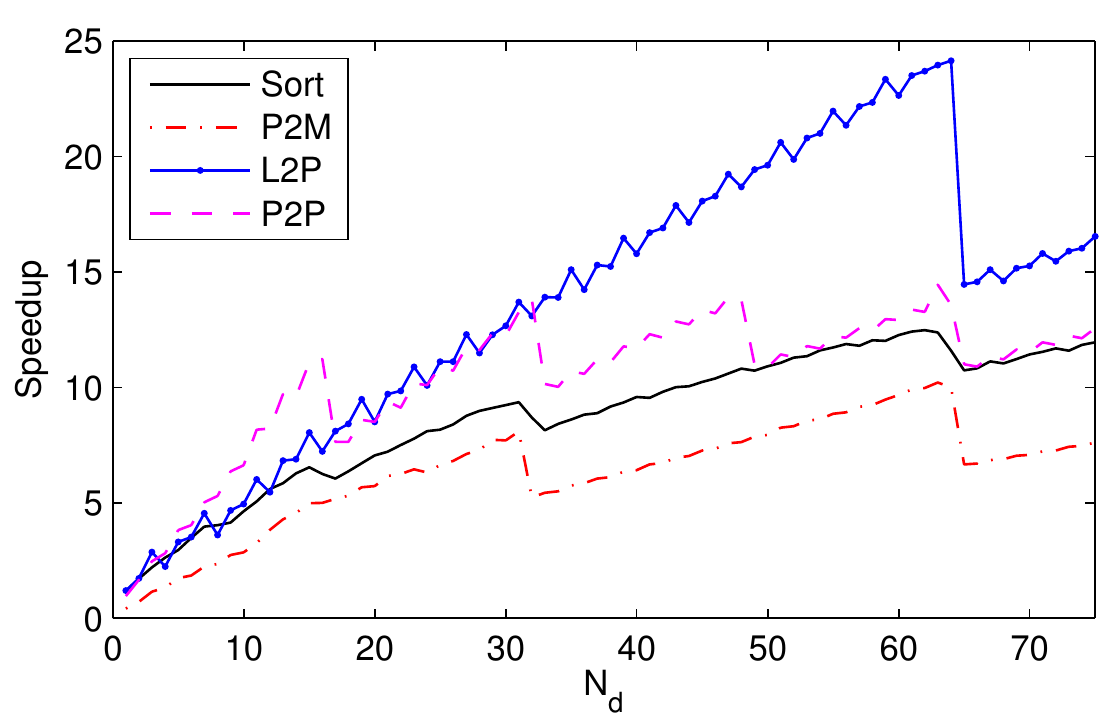}
  \caption{Speedup of the individual parts of the GPU-implementation
    as a function of the number of sources per box $N_{d}$. Here the
    total number of sources varies between $2^{16}$ and $75 \times
    2^{16}$.}
  \label{fig:Ndirspeedup}
\end{figure}

In Figure~\ref{fig:Ndirspeedup}, the GPU speedup as a function of the
number of sources per box is studied. Within this range, the shift
operators and the connectivity mainly depend on the number of levels
and therefore obtain constant speedups (hence we omit them). All parts
that depend on the number of particles in each box obtain higher
speedups for larger number of particles per box. This is expected,
since it is easier to get a good GPU load for larger systems. There is
also a performance decrease when the number of particles increases
above 32, 64, and so on, that is, at multiples of the warp size. The
direct evaluation additionally shows a performance decrease directly
after each multiple of 16, which is due to the fact that the algorithm
can use 4 threads per particle.

The small high frequency oscillations seen in the speedup of L2P and
P2P originates from the CPU algorithm, and is due to the use of SSE
instructions which makes the CPU code more efficient for an even
number of sources per box.

It should be noted again that the direct evaluation and connectivity
both make use of symmetry in the CPU version. This means that the
speedup would be significantly higher (almost a factor 2) if the CPU
version did not rely on this symmetry.

\begin{figure}
  \includegraphics{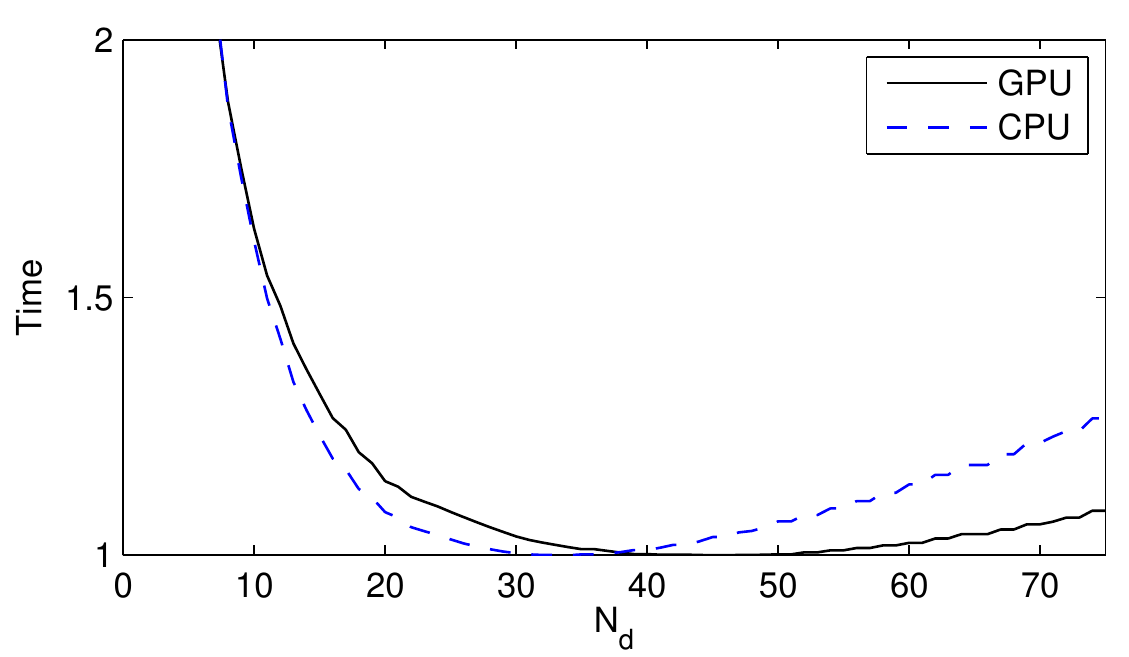}
  \caption{Time as a function of the number of sources per box $N_{d}$
    for the CPU and the GPU implementation, both normalized so that
    the fastest time $\equiv 1$.}
  \label{fig:Ndirect}
\end{figure}

As adding one extra level reduces the computational time of the direct
interaction, but increases the time requirement for most other parts,
it is necessary to find the best range of particles per box. The
number of levels $N_{l}$ is calculated according to
\begin{align}
  \label{eq:Nlevels}
  N_{l} &= \left\lceil 0.5\log_{2}
  \left( \frac{5}{8}\frac{N}{N_{d}} \right) \right\rceil,
\end{align}
where $N$ is the number of particles and $N_{d}$ is the desired number
of particles per box. This parameter choice was studied for 150
simulations with different number of particles (from $1 \times 10^4$
to $2 \times 10^6$). The result (normalized against the lowest time on
each platform) is found in Figure~\ref{fig:Ndirect}, showing that a
value around 45 is best for the GPU, while 35 is best for the CPU. Even
though the GPU has poor speedup for low number of particles, it still
scales better than the CPU in this case. The reason is that with low
values of $N_{d},$ the multipole shifts dominate the computational
time. This simulation was performed with 17 multipole coefficients,
giving a tolerance of approximately $1 \times 10^{-6}$. The tolerance
is here and below understood as
\begin{align}
  \TOL &= \left\Vert \frac{\Phi_{FMM}-\Phi_{exact}}{\Phi_{exact}}
  \right\Vert _{\infty}
\end{align}
where $\Phi_{exact}$ is the exact potential and $\Phi_{FMM}$ is the
FMM result.

\begin{table}
\centering
\begin{tabular}{lr|lr}
  Part & Time & Part & Time \\
  \hline 
  P2P & 43 \% & Connect & 1 \% \\
  Sort & 30 \% & M2M & $< 1$ \% \\
  M2L & 11 \% & L2L & $< 1$ \% \\
  P2M & 5 \% & & \\
  L2P & 2 \% &  Other & 8 \% \\
\end{tabular}

\caption{Time distribution of the GPU algorithm. The most expensive
  part in the case studied here is the direct evaluation (P2P),
  followed by sorting and M2L translations. The field \emph{``other''}
  contains all data transfers between the host and the GPU.}
\label{tab:pie}
\end{table}

For the optimal value 45 of $N_{d}$, the time distribution of the
different parts of the algorithm is given in Table~\ref{tab:pie} for
$N = 45 \times 2^{16}$, which gives 45 sources in each box at the
finest level of the FMM tree. According to \eqref{eq:Nlevels}, using
$N_{d} = 45$ gives 8 levels for $N \in (18 \times 2^{16},72\times
2^{16}]$. Within this interval, the time of P2P relative to the total
  time varies between 25\% to 55\%. It is particularly interesting to
  note in Table~\ref{tab:pie} that the sorting dominates by a factor
  of about 3 over the usually very demanding M2L-operation.

\subsection{Shift operators}

\begin{figure}
  \includegraphics{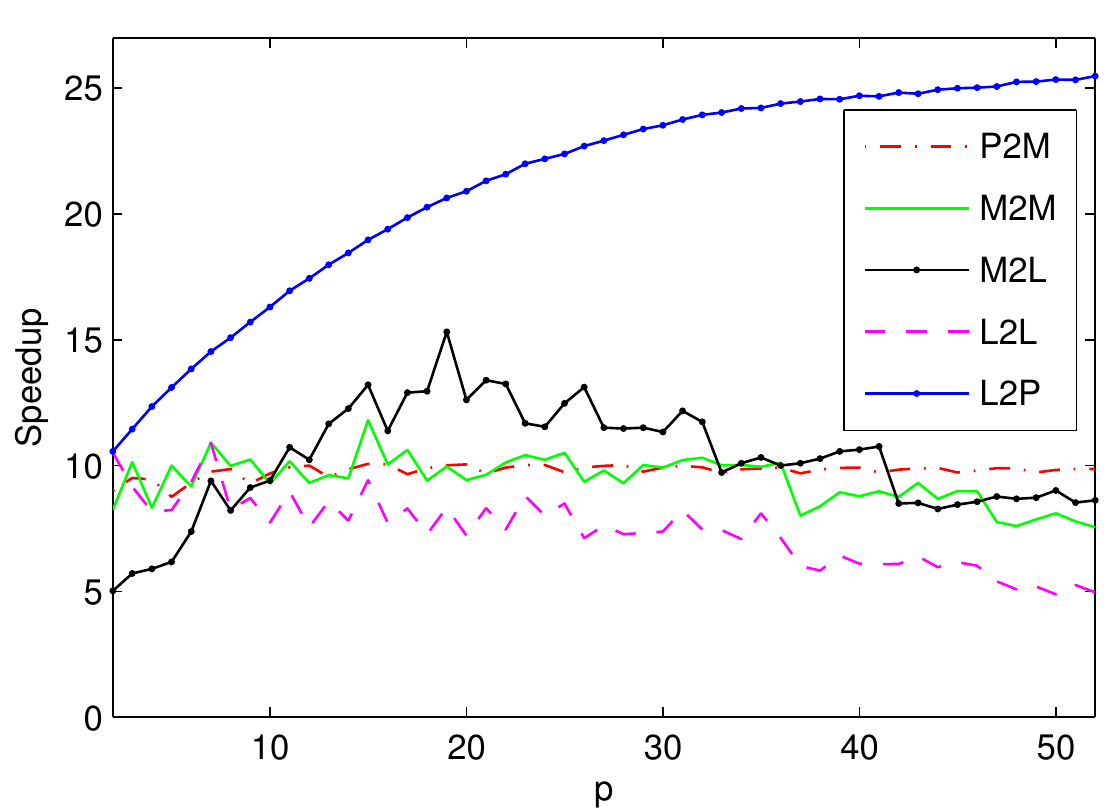}
  \caption{Speedup as a function of the number of multipole
    coefficients $p$.}
  \label{fig:pspeedup}
\end{figure}

The performance of the sorting and direct evaluation depends on the
number of sources per box and the number of levels while the
connectivity to a first approximation only depends on the number of
levels. The rest of the operators also depend on the number of
multipole coefficients (the number of multipole coefficients determine
the error in the algorithm). Multipole initialization and evaluation
depends linearly on the number of coefficients, while the shift
operators have two linear parts (pre- and post-scaling) and one
quadratic part (the actual shift). In the GPU case, all accesses to
global memory are included in the linear parts while all data is kept
in shared memory during the shift. A higher number of coefficients
increases the use of shared memory and fewer shift operations can
therefore be performed in parallel. The speedup as a function of
number of coefficients is plotted in Figure~\ref{fig:pspeedup}, where
the simulation was performed on $10^{6}$ particles with $N_{d} =
45$. The decrease in speedup due to lack of shared memory can be seen
quite clearly, e.g. at 42 coefficients for the M2L-shift, where one
block less (3 in total) can operate on the same multiprocessor.

The difference in speedup for L2P at low number of coefficients is
likely due to overhead, since these values stabilize at high enough
number of coefficients.

\begin{figure}
  \includegraphics{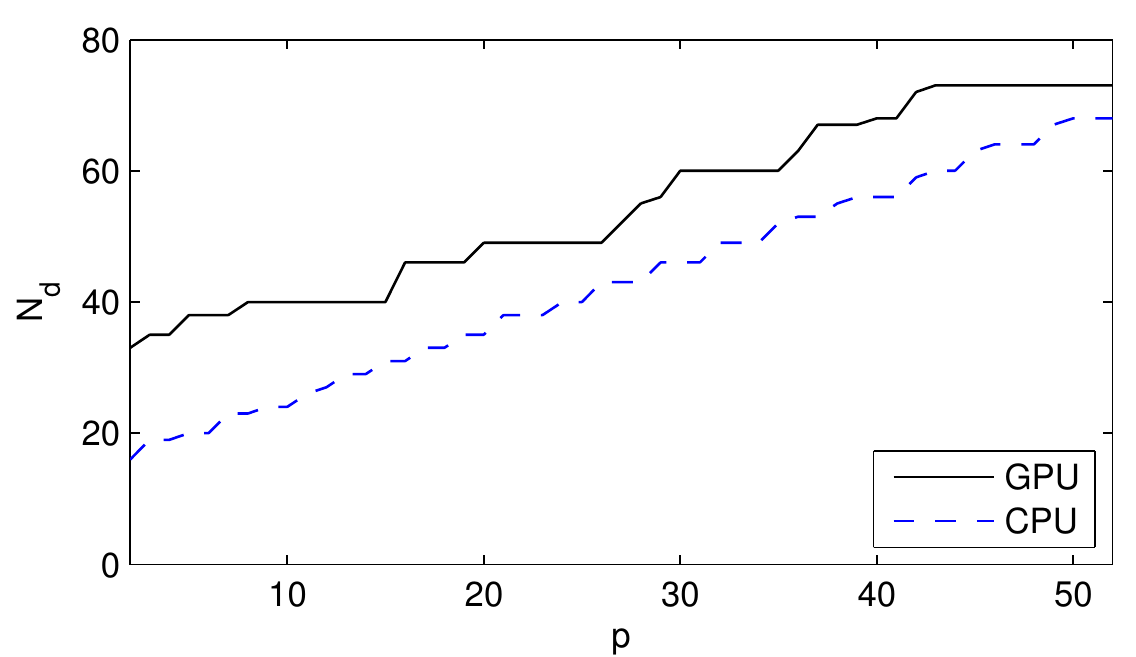}
  \caption{Optimal value of $N_{d}$ as a function of the number of
    multipole coefficients $p$.}
  \label{fig:Ndirecttol}
\end{figure}

Considering that the time required for the shift operators increases
with increasing number of coefficients, the optimal value for $N_{d}$
changes as well. Figure~\ref{fig:Ndirecttol} shows that the optimal
number for $N_{d}$ increases approximately linearly with increasing
number of coefficients.

\subsection{Break-even}
\label{subsec:breakeven}

\begin{figure}
  \includegraphics{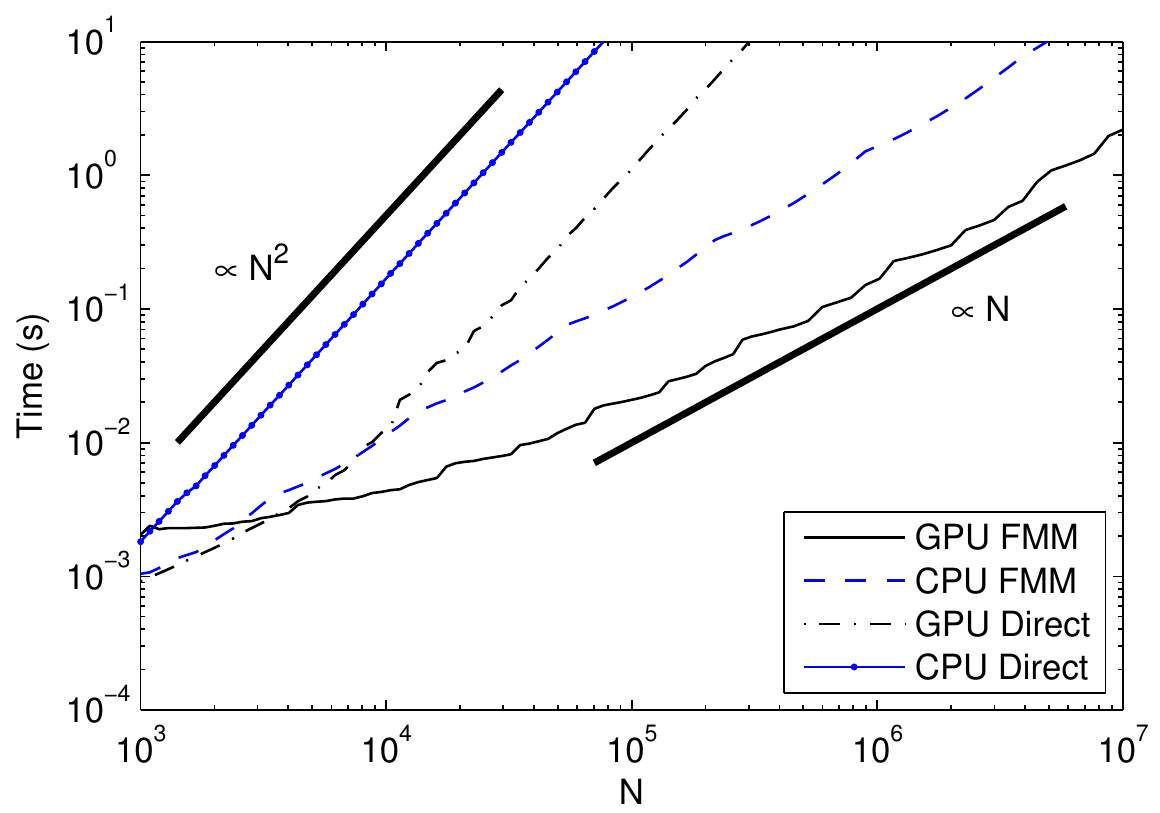}
  \caption{Total time of the algorithm as a function of the number of
    sources $N$. For the FMM-algorithm, the simulation was performed
    with $p = 17$, implying a tolerance of about $10^{-6}$.}
  \label{fig:breakeven}
\end{figure}

\begin{figure}
  \includegraphics{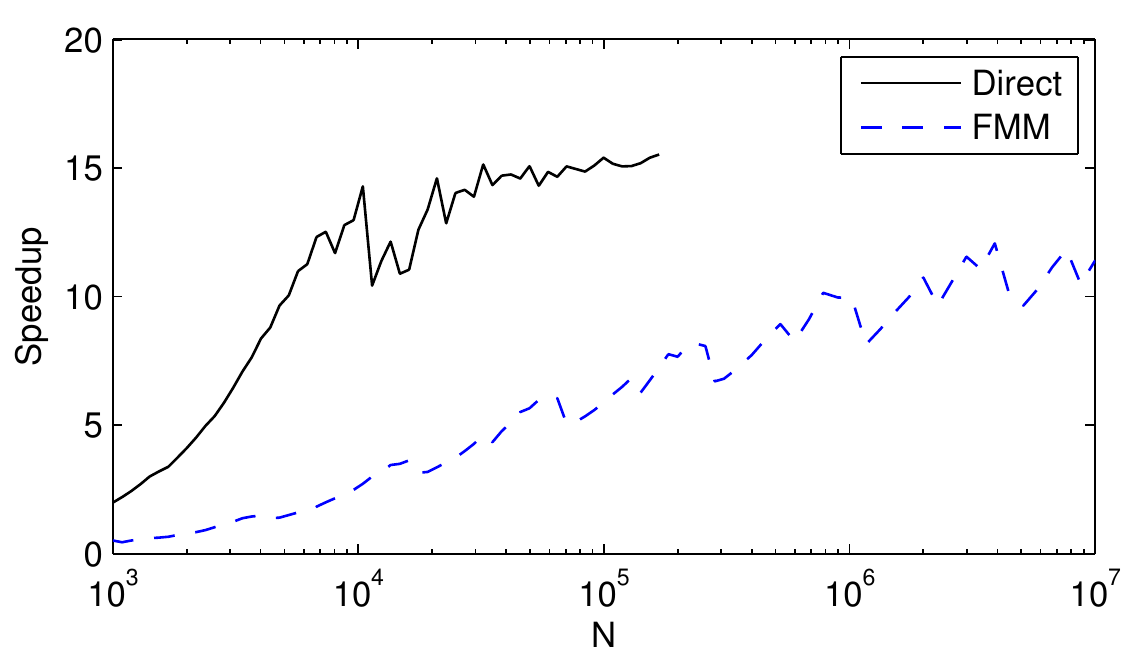}
  \caption{Speedup as a function of the number of sources $N$.}
  \label{fig:breakevenspeedup}
\end{figure}

If the number of sources is low enough, it may be faster to use a
direct summation instead of the fast multipole method. In
Figure~\ref{fig:breakeven}, the speed of the entire algorithm is
compared with the speed of direct summation for both the CPU and the
GPU implementation. The speedup of the GPU code increases with the
number of particles since more source points provide a better load for
the GPU. Looking at the direct summation times, the GPU scales
linearly in the beginning where the number of working cores is still
increasing, and later scales quadratically as the cores become fully
occupied. Since a double sum is easily performed in parallel and is
not memory bandwidth dominated the direct evaluation provides a good
estimate of the maximum speedup that can be achieved with the
GPU. Recall, however, that symmetry is used in the CPU implementation,
which almost speeds up the calculation with a factor of
2. Figure~\ref{fig:breakeven} shows that it is more beneficial to use
the FMM algorithm if the number of points exceeds about 3500 on the
GPU. This result compares favorably with that reported by others
\cite{fmmgpu3}. For large $N$, the speedup of the direct interaction
is higher than that of the FMM (15 compared to about 11, see
Figure~\ref{fig:breakevenspeedup}). Again, one should note that the
CPU version uses symmetry here. For simulations where the source
points and evaluation points are separate, the speedup is about 30 for
the direct evaluation and 15 for the FMM. The lower increase in
speedup for the FMM is due to the fact that only the P2P-evaluation of
the algorithm uses this symmetry (compare Table~\ref{tab:pie}).

\begin{figure}
  \includegraphics{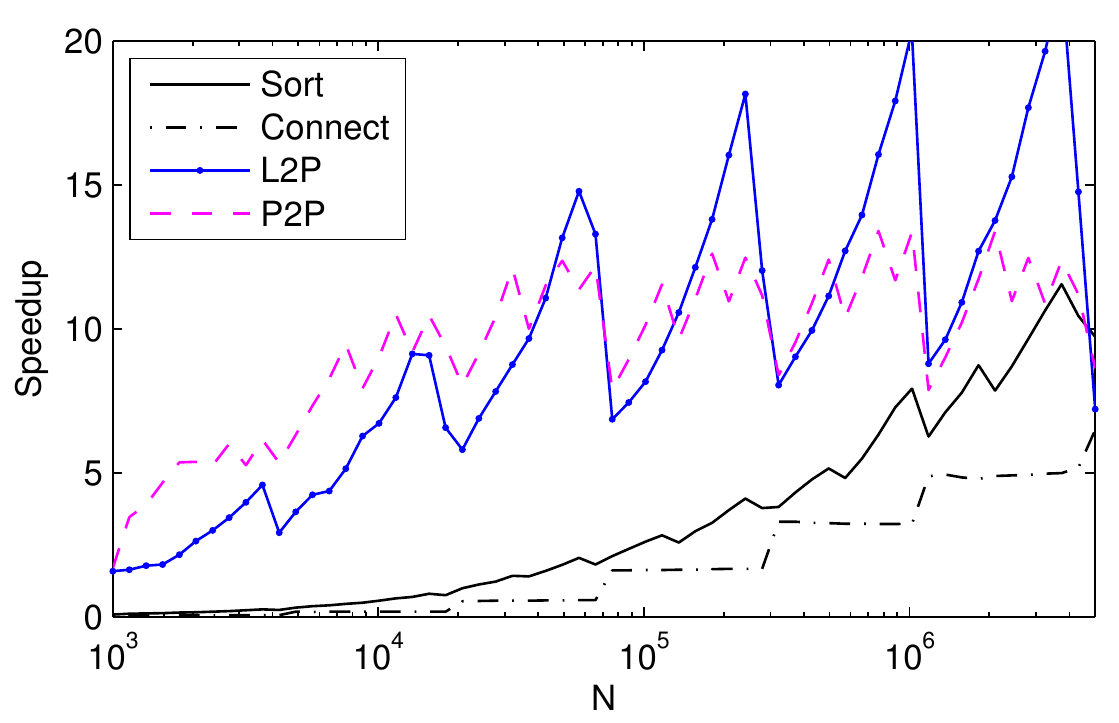}
  \includegraphics{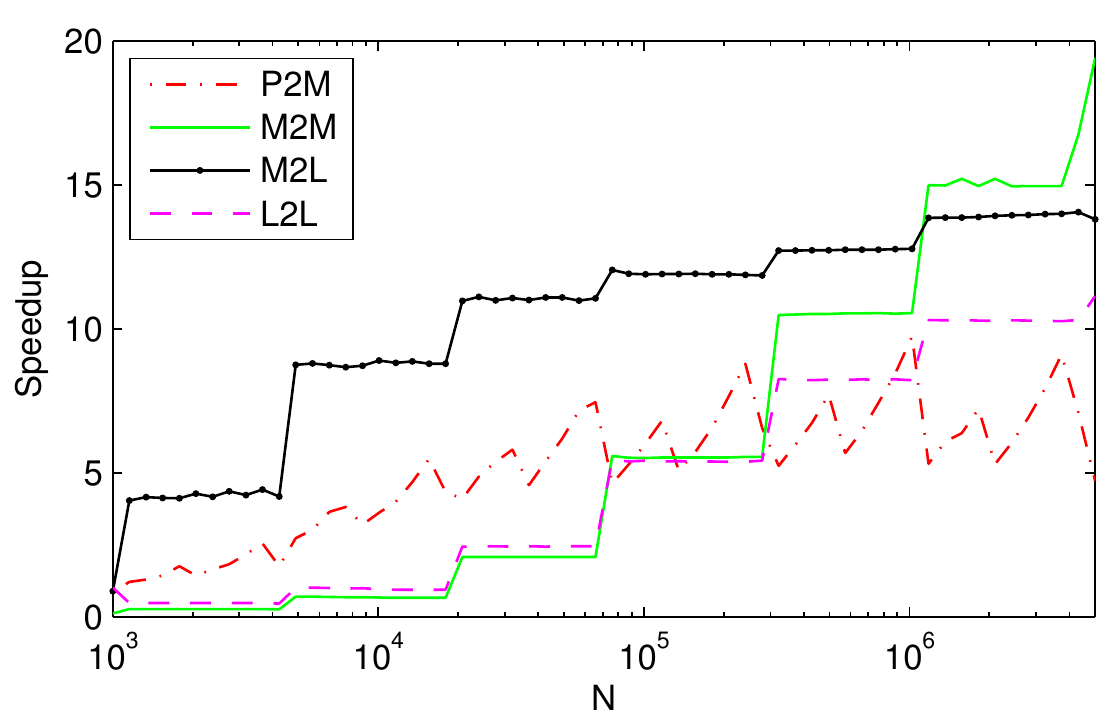}
  \caption{Speedup of individual parts as a function of the number of
    sources.}
  \label{fig:Nspeedup}
\end{figure}

Comparing the individual parts (Figure~\ref{fig:Nspeedup}), the M2L-
and P2P-shifts quite rapidly obtain high speedups, while the sorting
requires quite a large number of points. The poor values for M2M and
L2L at low number of particles are due to the fact that few shifts are
performed at the lower levels, causing many idle GPU cores. The
situation is the same for the connectivity. As these algorithms have
to be performed one level at a time, the low performance of the shifts
high up in the multipole tree decreases the performance of the entire
step. Consequently, the speedup increases with an increasing number of
source points. The oscillating behavior of the multipole
initialization and evaluation is related to the number of particles in
each box (compare with Figure~\ref{fig:Ndirspeedup}).

The code has been tested both on the Tesla platform used for the above
figures, and on a Geforce GTX 480 platform (which has 480 cores,
compared to 448 for the Tesla card). The total run-time is
approximately the same on both platforms. Notable differences are that
the P2P interaction is faster on the Tesla if $N_d$ is high, and in
the simulation in Table~\ref{tab:pie}, the GTX 480 card required 30\%
longer time than the Tesla card. On the other hand, the Tesla card
required 25\% longer time than the GTX 480 for the sorting (which is
limited by memory access, rather than double precision math). The
shift operators were approximately equally fast on both systems. The
overall result is that the optimal value for $N_d$ is lower for the
GTX 480 card (35 instead of 45) for a total running time which was
approximately the same. This shows that the much cheaper GTX 480
gaming card is a perfectly reasonable option for this implementation
of the fast multipole method, despite the fact that it is written in
double precision.

\subsection{Benefits of adaptivity}

\begin{figure}
  \includegraphics{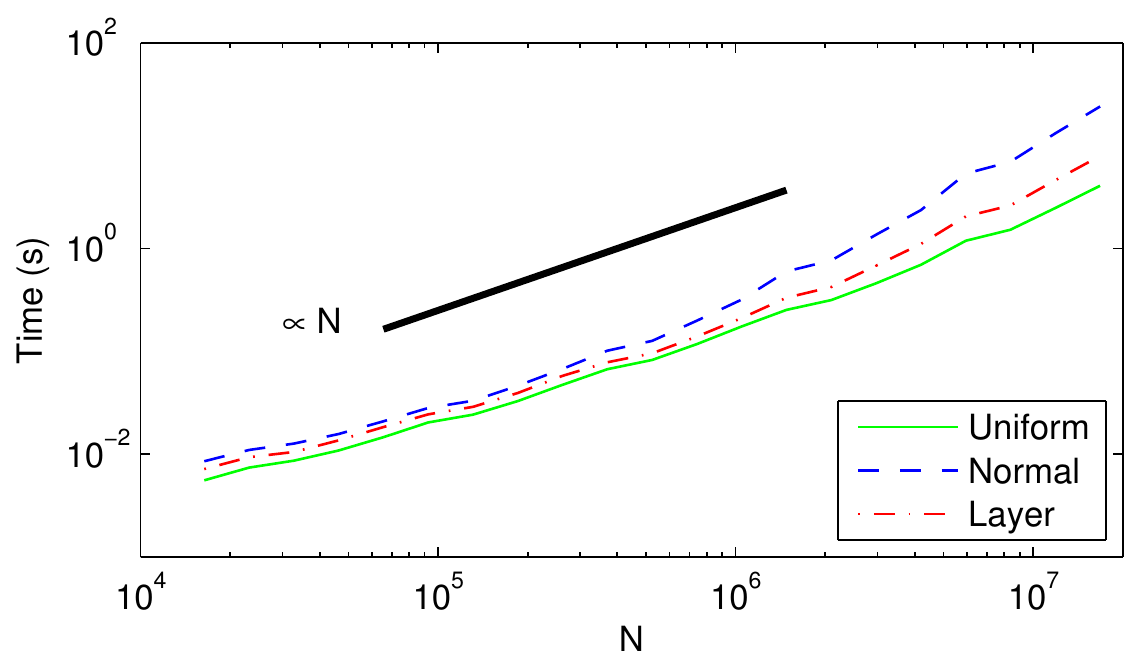}
  \caption{Performance of the code when evaluating the harmonic
    potential for three different distributions of points. The source
    points were \textit{(i)} uniformly distributed in $[0,1] \times
    [0,1]$, \textit{(ii)} normally distributed with variance $1/100$,
    and \textit{(iii)} distributed in a `layer' where the
    $x$-coordinate is uniform, and the $y$-coordinate is again
    $N(0,1/100)$-distributed. For the purpose of comparison, all
    distributions were rejected to fit exactly within the unit
    square. The FMM mesh for case \textit{(ii)} is shown in
    Figure~\ref{fig:mesh}.}
  \label{fig:adaptivity}
\end{figure}

As a final computational experiment we investigated the performance of
the adaptivity by using different point distributions. Under a
relative tolerance of $10^{-6}$ ($p = 17$ in \eqref{eq:multipole} and
\eqref{eq:local}) we measured the evaluation time for increasing
number of points sampled from three very different distributions. As
shown in Figure~\ref{fig:adaptivity}, the code is robust under highly
non-uniform inputs and scales well at least up to some 10 million
source points.

\begin{figure}
  \includegraphics{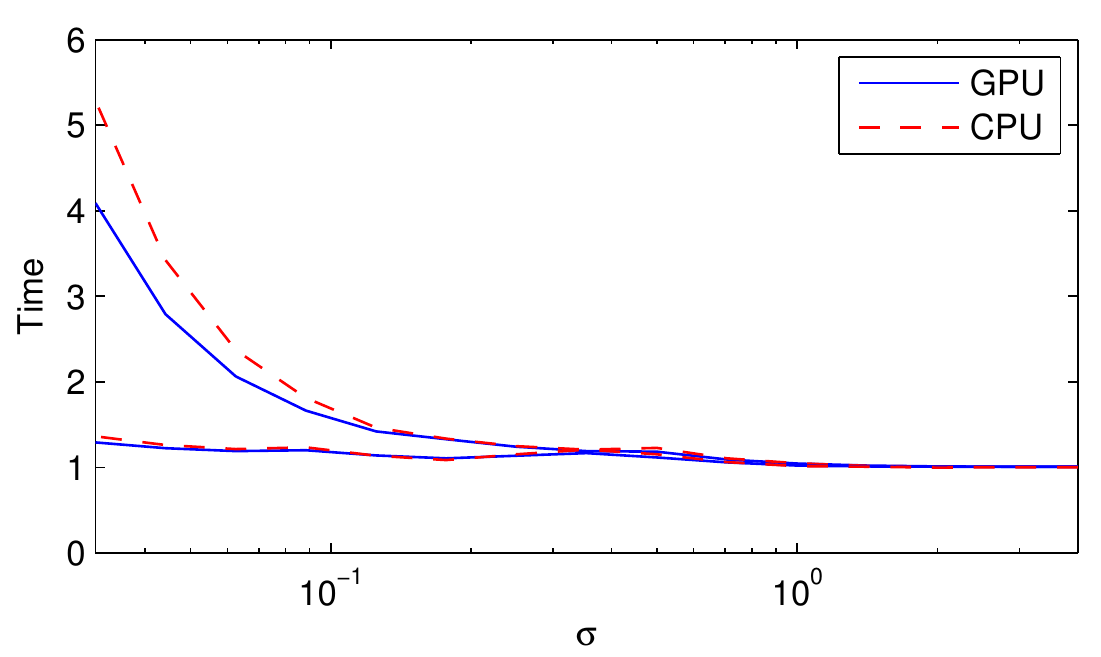}
  \caption{Robustness of adaptivity. Time for two different
    non-uniform distributions normalized with respect to a uniform
    distribution of points. The top two graphs are for the normal
    distribution of sources, while the lower two graphs are for the
    `layer' distribution. See text for further details.}
  \label{fig:nonuniform}
\end{figure}

When the distribution of particles is increasingly non-uniform, more
boxes will be in each others near-field resulting in more direct
interactions. This is tested in Figure~\ref{fig:nonuniform} for the
two non-uniform distributions from Figure~\ref{fig:adaptivity}. Both
the CPU- and GPU timings have been normalized to the time of a
homogeneous distribution. The results indicate that the decrease in
performance for highly non-uniform distributions is less for the GPU
version than for the CPU version. From the timings of the individual
parts, it is seen that almost all the increase in computational time
originates from the P2P-shift. The speedup for all time consuming
individual parts is relatively constant with respect to the degree of
non-uniformity (e.g. the parameter $\sigma$ in
Figure~\ref{fig:nonuniform}) and the reason the GPU code handles a
highly non-uniform distribution better is simply because the P2P
evaluation has a higher speedup than the overall code.

\section{Conclusions}
\label{sec:conclusions}

We have demonstrated that all parts of the adaptive fast multipole
algorithm can be efficiently implemented on a GPU platform in double
precision. Overall, we obtain a speedup of about a factor of 11
compared to a highly optimized (including SSE intrinsics), albeit
serial, CPU-implementation. This factor can be compared with the
speedup of about 15 which we obtain obtained for the direct $N$-body
evaluation, a task for which GPUs are generally understood to be well
suited \cite{moldynGPU} (see Figure~\ref{fig:breakevenspeedup}).

In our implementation, all parts of the algorithm achieve speedups in
about the same order of magnitude. Generally, we obtain a higher
speedup whenever the boxes contain some 20--25\% more particles than
the CPU version (see Figures~\ref{fig:Ndirecttol} and
\ref{fig:Nspeedup}). Given the data-parallel signature of this
particular operation, this result is quite expected. Another
noteworthy result is that our version of the GPU FMM is faster than
the direct $N$-body evaluation at around $N = 3500$ source points, see
Figure~\ref{fig:breakeven}. Our tests also show that the asymmetric
adaptivity works at least as well on the GPU as on the CPU, and that
in some cases it even performs better.

When it comes to \emph{coding complexity} it is not so straightforward
to present actual figures, but some observations at least deserve to
be mentioned. The topological phase was by far the most difficult part
to implement on the GPU. In fact, the number of lines of codes
approximately \emph{quadrupled} when transferring this operation. We
remark also that the topological phase performs rather well on the
GPU, an observation which can be attributed to its comparably high
internal bandwidth. Thus, there is a performance/software complexity
issue here, and striking the right balance is not easy.

By contrast, the easiest part to transfer was the direct evaluation
(P2P), where, due to SSE-intrinsics, the CPU-code is in fact about
twice the size than the corresponding CUDA-implementation.

These observations as well as our experimental results, suggest that a
\emph{balanced} implementation, where parts of the algorithm are
off-loaded to the GPU while the remaining parts are parallelized over
the CPU-cores, would be a reasonable compromise. This has also been
noted by others \cite{BarnesHutGPU} and is ongoing research.





\subsection{Reproducibility}
\label{subsec:reproduce}

Our implementation as described in this paper is available for
download via the second author's
web-page\footnote{\url{http://user.it.uu.se/~stefane/freeware}}. The
code compiles both in a serial CPU-version and in a GPU-specific
version and comes with a convenient Matlab mex-interface. Along with
the code, automatic Matlab-scripts that repeat the numerical
experiments presented here are also distributed.


\section*{Acknowledgment}

We like to acknowledge inputs from the UPMARC@TDB research group and
Olov Ågren.

This work was financially supported by the Swedish Energy Agency,
Statkraft AS, Vinnova and Uppsala University within the Swedish Centre
for Renewable Electric Energy Conversion (A.~Goude), and the Swedish
Research Council within the UPMARC Linnaeus center of Excellence
(S.~Engblom).


\newcommand{\doi}[1]{\href{http://dx.doi.org/#1}{doi:#1}}
\newcommand{\available}[1]{Available at \url{#1}}
\newcommand{\availablet}[2]{Available at \href{#1}{#2}}


\end{document}